\documentclass[sigconf,nonacm]{acmart}
\usepackage{algorithm}
\usepackage{algpseudocode}
\usepackage{caption}
\usepackage{multirow}

\AtBeginDocument{%
  \providecommand\BibTeX{{%
    \normalfont B\kern-0.5em{\scshape i\kern-0.25em b}\kern-0.8em\TeX}}}

\begin{document}

\title{Learning Embedded Representation of the Stock Correlation Matrix using Graph Machine Learning}

\author{Bhaskarjit Sarmah}
\email{bhaskarjit.sarmah@blackrock.com}
\affiliation{%
  \institution{BlackRock Inc.}
  \city{Gurgaon}
  \country{India}
}

\author{Nayana Nair}
\email{nayana.nair@blackrock.com}
\affiliation{%
  \institution{BlackRock Inc.}
  \city{Gurgaon}
  \country{India}
}

\author{Dhagash Mehta}
\email{dhagash.mehta@blackrock.com}
\affiliation{%
  \institution{BlackRock Inc.}
  \city{New York}
  \country{USA}
}

\author{Stefano Pasquali}
\email{stefano.pasquali@blackrock.com}
\affiliation{%
  \institution{BlackRock Inc.}
  \city{New York}
  \country{USA}
}





\begin{abstract}
  Understanding non-linear relationships among financial instruments has various applications in investment processes ranging from risk management, portfolio construction and trading strategies. Here, we focus on interconnectedness among stocks based on their correlation matrix which we represent as a network with the nodes representing individual stocks and the weighted links between pairs of nodes representing the corresponding pair-wise correlation coefficients. The traditional network science techniques, which are extensively utilized in financial literature, require hand-crafted features such as centrality measures to understand such correlation networks. However, manually enlisting all such hand-crafted features may quickly turn out to be a daunting task. Instead, we propose a new approach for studying nuances and relationships within the correlation network in an algorithmic way using a graph machine learning algorithm called Node2Vec. In particular, the algorithm compresses the network into a lower dimensional continuous space, called an embedding, where pairs of nodes that are identified as similar by the algorithm are placed closer to each other. By using log returns of S\&P 500 stock data, we show that our proposed algorithm can learn such an embedding from its correlation network. We define various domain specific quantitative (and objective) and qualitative metrics that are inspired by metrics used in the field of Natural Language Processing (NLP) to evaluate the embeddings in order to identify the optimal one. Further, we discuss various applications of the embeddings in investment management.
\end{abstract}


\maketitle

\section{Introduction}
In the financial domain, it is not only intuitively appreciated that financial systems are intricately connected with each other but the interconnectedness among financial entities such as assets, banks, managers (fund managers as well as higher management of companies), etc. have also been extensively investigated using network science approaches\cite{allen2009networks,saha2022survey}. In such a financial network, the nodes represent the financial entities and links between pairs of nodes represent a well-defined relation such as cash flows, holdings of shares, or financial exposures between them. The financial networks may come in various flavors ranging from regular unweighted, undirected, weighted, directed, bi-directed, bipartite, multi-graph, multiplex, bipartite, dynamic (time-dependent), etc.\cite{bardoscia2021physics}.

The applications of network science in asset management industry include understanding the effect of a potential price fluctuation in a specific sector or stock price on a fund universe where all the funds and their underlying holdings form a bipartite network; modeling systematic risk and risk propagation among funds, for example, to understand how selling a subset of assets from one of the funds affects the liquidity of other portfolios; in constructing various portfolio diversification measures for a fund or for a fund of funds; stock selection to construct diversified portfolios, etc. (see, e.g., \cite{wu2015centrality, banerjee2013diffusion, huang2016financial,onnela2003dynamics,pozzi2013spread,acemoglu2015systemic,peralta2016network, delpini2019systemic})

Of particular interest to the present work is networks of correlations among financial assets. Correlation matrices have been one of the most studied objects in finance as their role in portfolio construction \cite{10.2307/2975974}, in capital asset pricing model \cite{fama2004capital}, factor analysis \cite{fabozzi2014basics}, etc. Correlation matrices of financial assets, especially stock correlation matrices, where each node corresponds to an individual stock and the link between each pair of stocks is the corresponding correlation, have been extensively investigated in the literature \cite{mantegna1999hierarchical,bonanno2003topology,bonanno2004networks,peralta2016network,PhysRevX.5.021006,aste2010correlation}.

Most of the aforementioned research, however, primarily relies on the computation of statistical properties \cite{newman2018networks} of networks such as degree centrality, closeness centrality, eigenvector centrality, average shortest path, clustering coefficient, to name a few. Network properties such as these arrive in various flavors and can be computed using a weighted or unweighted network, a directed or undirected network and so on. In such cases, it can become a manual exercise for a researcher to create features that capture relationships between the nodes of a network which can be further used in downstream applications.

\subsection{Our Contributions and Previous Works}
In the present work, we propose a machine learning (ML) based algorithm to learn a low-dimensional representation, called embedding, of the stock correlation network. Learning embeddings is a common task in Natural Language Processing (NLP) where words are represented by vectors in an abstract low dimensional manifold \cite{mikolov2013distributed, pennington2014glove}. Here, semantically similar words are identified from the given corpus of text and placed together in the manifold. We then use correlation matrix of log returns of the S\&P 500 stocks, represent it as a network and then apply Node2Vec \cite{grover2016node2vec} to create sentence like structures (i.e., directed subnetworks) by generating multiple random walks from each node. Then, we use a word embedding algorithm called Word2Vec \cite{mikolov2013distributed} to learn the embedding from the generated data. We also propose several domain specific metrics to directly evaluate the embeddings rather than through downstream tasks unlike most other works in this area: we evaluated the stock embeddings by tuning hyperparamters by optimizing the \emph{V-measure} between the clusters in the embedding and Global Industry Classification Standard (GICS) categories. We also used stock similarity and analogical inference of stocks as further qualitative evaluation metrics of the embeddings. 

There has been extensive research making use of graph ML to learn stock embeddings in the literature \cite{saha2022survey}. However, in all the existing research, the respective embeddings are learned by tuning hyperparameters based on a downstream task. e.g., one of the closest works to the present work is Ref.~\cite{saha2021stock} where the authors optimized stock embeddings to rank stocks according to the predicted return by using Node2Vec where the hyperparameters of Node2Vec were tuned on a downstream stock ranking prediction task. Moreover, our goal is not to rank stocks rather to identify similarity.

Ref. \cite{long2020integrated} have used a deep neural network framework to predict stock price trends using transaction records and public market information. Ref.~\cite{wang2020stock2vec} whose goal was to predict stock prices learned stock embeddings as a by-product of a trained temporal convolutional network using time-series data for S\&P 500 stocks. Since they posed it as a supervised forecasting task, the hyperparameters were tuned using regression metrics such as root mean squared error and mean absolute percentage error. Ref.~\cite{yi2022stock2vec} used S\&P 500 stock price as well as daily trading information including trading date, opening price, highest price, lowest price, closing price, the number of shares traded, and company stock names for computing stock embeddings using Word2Vec and evaluated the embeddings based on their performance on a downstream task, i.e., by using Word2Vec embeddings on four classifiers (Gaussian Naive Bayes, Support Vector Machines, Decision Tree and the Random Forest). Then, if one of these classifiers achieves a higher accuracy to predict the stocks' industry sectors with the Word2Vec embedding at a given dimensionality, they reasoned that the Word2Vec embedding with that dimensionality can better represent the original data. Along the same lines, and again closer to the present work, Ref.~\cite{dolphin2022stock} constructed context-target stocks data based on closest stocks in terms of their returns to the target stocks before training stock embeddings using Word2Vec. Then, they used a classification model with embeddings as input and industry sector as the output to evaluate embeddings.

In our approach, we do not rely on another classifier on top of our embedding unlike \cite{yi2022stock2vec}, and we also keep the learning task as an unsupervised one unlike \cite{saha2021stock}.

Some of the other relevant works which have attempted to learn stock embeddings but use alternative data are as follows: in Ref.~\cite{lu2021stock}, stock embeddings were computed using stock news and sentiment dictionaries to predict stock trends. Ref.~\cite{wu2019deep} utilized co-occurence matrix of stocks mentioned in news articles and truncated singular value decomposition approach from GloVe algorithm \cite{pennington2014glove} to compute stock embeddings.

To summarize, in addition to a novel way of viewing the correlation network as text data and then applying word embedding model, the present work proposes an objective metric to directly evaluate embeddings and in turn tune hyperparameters based on three increasingly granular levels of GICS classifications. Once the embedding is learned at an optimal hyperparameter point, the embedding can then be used to determine stock similarity as well as for analogical inference among stocks which are directly \textit{anchored to the GICS classifications}. 


\section{Data Preprocessing and Network Statistics}
For the purpose of this work, we rely on the publicly available returns data for S\&P 500 stocks for the year 2021. In this Section, we provide details of data preprocessing and the process of constructing a filtered network out of the correlation matrix.

\subsection{Data Preprocessing}
\subsubsection{Data Cleaning}
We web scrapped S\&P 500 wikipedia page to get a list of tickers that constitutes S\&P 500 index. We used yahoo finance, a publicly available data source, to collect price data for these tickers. Price data is collected for the entire year of 2021 which includes 504 common stocks issued by 500 companies. We used end of day adjusted close prices of these stocks for this analysis. There are no missing data in adjusted close price as no stock was added in or removed from the index in 2021.

\subsubsection{Building a Network of Stocks}
To create an undirected weighted network for the S\&P 500 universe, we use a widely adopted technique popularized in Ref.~\cite{mantegna1999hierarchical}:
each node of the network represents a stock, the link between the nodes corresponds to whether the pair of stocks is correlated to each other, and the weight on the link corresponds to the actual correlation of the log returns of the pair of stocks. Here, for the $i$-th stock, the log returns are calculated using daily adjusted close prices as
\begin{displaymath}
    r_{i}(t)=\log P_{i}(t)-\log P_{i}(t-1),
\end{displaymath}
where $P_{i}(t)$ denotes the daily closing price of the $i$-th stock at the $t$-th day and $r_{i}(t)$ denotes the return of the $i$-th stock at the $t$-th day.

\begin{figure}[ht]
    \centering
    \includegraphics[width=8cm, height=8cm]{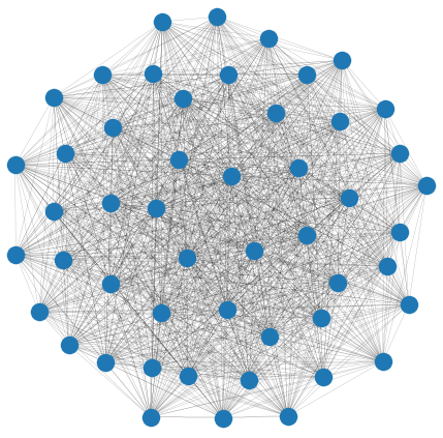}
    \caption{Network of correlations of S\&P 500 stocks based on their log returns. Here, each node is a stock and the edge between them is weighted as $\rho_{i j}$, hence the network is a weighted fully connected network.}
    \label{fig:1}
\end{figure}

Then, the correlation, $\rho_{i j}$, between the $i$-th and $j$-th stocks is computed as
\begin{displaymath}
    \rho_{i j}=\frac{\left\langle r_{i} r_{j}\right\rangle-\left\langle r_{i}\right\rangle\left\langle r_{j}\right\rangle}{\sqrt{\left(\left\langle r_{i}^{2}\right\rangle-\left\langle r_{i}\right\rangle^{2}\right)\left(\left\langle r_{j}^{2}\right\rangle-\left\langle r_{j}\right\rangle^{2}\right)}}.
\end{displaymath}
Here, $i$ and $j$ runs from $1, \dots, n$, where $n$ is the total number of stocks. From the correlation matrix $\rho$ one can construct a fully connected network from the correlation matrix as depicted in Figure \ref{fig:1}.
Ideally, analysing the weighted fully connected network may provide complete information about the underlying relationship among the stocks. However, traditionally most of the network quantities are analyzed for sparse networks as most real world networks are sparse \cite{newman2018networks}. Moreover, computational complexities to compute network quantities increases for denser networks and may be the worst for complete networks. Hence, the next step in the present work is to algorithmically sparsify the complete network, however we plan to investigate complete networks in the future.

To sparsify a given weighted fully connected network by identifying and removing `unimportant' links, there are various algorithms that can be used such as Minimum Spanning Tree (MST) \cite{mantegna1999hierarchical}, Average Linkage Minimum Spanning Tree \cite{tumminello2007spanning}, Planar Maximally Filtered Graph \cite{tumminello2005tool}, etc. Traditionally, a popular choice to learn correlation networks in finance is MST. However, MST works in an inverse way than preferred in our case: for a weighted complete network, MST preserves those links of the network which can be used to traverse the entire network with minimum distance, whereas in the stock correlation network the links with low correlations are intuitively less important and hence should be removed. 

\begin{figure}[ht]
    \centering
    \includegraphics[width=8cm, height=8cm]{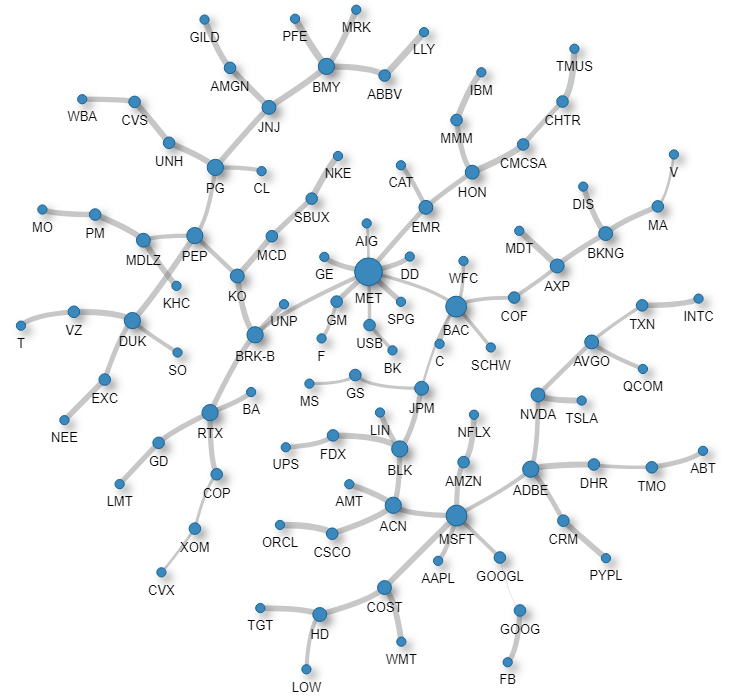}
    \caption{Network after applying MST algorithm.}
    \label{fig:2}
\end{figure}

To resolve this technical issue, i.e., to filter the fully-connected graph while preserving links with high correlation, we first convert the correlation matrix to a distance matrix where distance between pair of stocks $i$ and $j$ is computed as
\begin{displaymath}
d_{ij} = \sqrt{2(1 - \rho_{ij})}.
\end{displaymath}
Applying MST on the distance matrix, $d$, removes edges which have high distance (low correlation). Then the edges of the filtered distance network are replaced with their corresponding correlation coefficients to obtain a filtered correlation network. Figure \ref{fig:2} shows the sparsified network of correlations of log returns of S\&P 500 stocks using MST.

In summary, after applying MST algorithm we obtain a sparse weighted network of S\&P 500 stocks which we can begin to analyze using network science techniques.

\subsection{Network Statistics}
Here, we provide details of certain network quantities as a warm up to the graph machine learning based analysis. We start by providing basic summary statistics for the filtered network.
\subsubsection{Computational Details}
In the present work, we have used NetworkX \cite{hagberg2008exploring}
library in Python to compute network statistics and apply MST on the stock network and 
StellarGraph \cite{StellarGraph} and Gensim \cite{vrehuuvrek2011gensim} library to apply Node2Vec and Word2Vec respectively.

\subsubsection{Basic Network Statistics}
Before delving into the ML algorithm, we briefly provide details on the basic statistics. The statistics are summarized in Table \ref{tab:unweighted_network_statistics} for the final filtered weighted network. Here, the degree of a node is the number of links from a node, the shortest path between a pair of nodes is the path on the network between the given pair of nodes that has the shortest weighted path-length, and the diameter of the network is the longest shortest path-length in the network.

\begin{table}[]
\begin{tabular}{l l l}
\toprule
Unweighted Quantity  &  Value  \\ \hline
\hline
Number of Nodes & 504   \\
Number of Links & 503   \\
Average Degree & 2   \\
Diameter of the network & 36   \\
Average shortest path & 13.97   \\
\bottomrule
\end{tabular}
\caption{Unweighted network quantities for the stock network.}\label{tab:unweighted_network_statistics}
\end{table}

\section{Methodology}
One can resort to the above weighted network quantities to compute similarity among nodes of a network. However, manually enlisting all possible network quantities is a prohibitively difficult task. Instead, we propose an ML based approach to learn an embedded representation of the network data such that similar nodes are placed together in a lower dimensional manifold. Such a representation is learned directly from the raw network data rather than non-linearity supplied explicitly. In the following, we will also describe additional advantages of learning such a representation. Below, however, we begin by explaining the specific algorithm used in the present work, called Node2Vec.

Node2Vec \cite{grover2016node2vec} is one of the most popular algorithms used to learn lower dimensional representation for nodes in graph. In Node2Vec, we learn a mapping of nodes to a low-dimensional space of features that maximizes the likelihood of preserving the network neighborhoods of nodes. In this paper, we use Node2Vec to create sentence-like structures from a given weighted network using a second-order random biased walk. Then, we construct a set of context stocks and target stocks using a moving window using the Word2Vec algorithm. Below we briefly explain both Word2Vec and Node2Vec.

\subsection{Embedding Learning}
We begin to explain our ML methodology by first briefly describing the Word2Vec algorithm. Word2Vec was developed in Ref.~\cite{mikolov2013distributed} and is one of the most widely used word embedding techniques in natural language processing (NLP). Traditional approaches of feature extraction such as bag-of-words, tf-idf, one-hot-encoding, do not capture the semantic similarities among words. Word2Vec overcomes this problem by constructing a lower dimensional representation that captures meaningful semantic and syntactic relationships between words in the corpus. Here, one trains a shallow neural network to predict the target word provided a set of context words by minimizing categorical cross entropy loss. Once the neural network is trained, word embeddings are computed as average of the weight matrices obtained from the hidden layer and the output layer, respectively. Here, the word embeddings are obtained as a by-product of the model training process where the target word is predicted for a set of context words. Word2Vec uses one of the two architectures, namely \emph{Continuous Bag-of-Words (CBOW)} (where the surrounding context words are used to predict the target word) or \emph{Skip-gram} (where the target word is fed as input and the context or surrounding words is generated as the output) model to create word embeddings.

Since a corpus of sentence-like structures is required to learn stock embeddings using Word2Vec, to generate such structures from the available static weighted network of stock correlations, we make use
of the Node2Vec \cite{grover2016node2vec} algorithm: the algorithm samples a set of random walks of specific length starting from each node of the given network. The set of nodes in the network is considered as the \emph{vocabulary} and the directed path of each random walk is considered as a \emph{sentence}. This corpus of sentence-like structures can then be fed into Word2Vec to obtain the desired embedding. In short, stock embeddings are then learned using Word2Vec algorithm applied to the set of weighted biased random walks performed over the network (see Table \ref{tab:context_target_stocks} for an example dataset to be fed into Word2Vec).

\begin{table}[ht]
    \centering
   \begin{tabular}{|c|c|}
    \hline
    \bfseries Context Stocks & \bfseries Target Stock\\ \hline
        A, ALGN, NOW, ETR & TGT  \\ \hline
        ALGN, TGT, ETR, ROL& NOW \\ \hline
        IR, NSC, GNRC, HOLX & GPN \\ \hline
        NSC, GPN, HOLX, PTC & GNRC \\ \hline
        A, ICE, SYF, WAB & OKE \\ \hline 
        ICE, OKE, WAB, DG & SYF \\ \hline
    \end{tabular}\caption{An example of training data generated using Node2Vec out of the filtered network to be fed into the Word2Vec algorithm.}\label{tab:context_target_stocks}
    \vspace{-2em}
\end{table}

\subsection{Hyperparameters}\label{sec:hyperparameters}
It has been shown that tuning of the hyperparameters for both Node2Vec\cite{hacker2022surprising} and Word2Vec \cite{caselles2018word2vec} is crucial to obtain robust embeddings.

The following hyperparameters have to be tuned in order to create the required corpus of sentence-like structures from the filtered network using Node2Vec: the number of random walks $r$ from each node in the network; the length for each random walk $l$ from each node in the network; the probability $p$ with which a random walk will return to the node it visited previously; and, the probability $q$ with which a random walk will explore the unexplored part of the graph.

Word2Vec algorithm has the following hyperparameters which have to be tuned in order to quantitatively evaluate the strength of the embeddings generated:
\begin{enumerate}
    \item Window size, $w$: maximum distance between the current and predicted word within a sentence;
    \item Vector size, \emph{dim}: the dimensionality of the word vectors.
\end{enumerate}

\section{Evaluation Metrics}
By definition, there may not be any objective metrics to evaluate or compare the final results when the learning task is performed in a genuinely unsupervised fashion such as the problem at hand \cite{mara2019evalne,pellegrino2020geval}. 

To come up with metrics to evaluate embeddings for our problem, we resort to NLP for inspiration. Refs.~\cite{bakarov2018survey,wang2019evaluating} categorize various metrics used to evaluate word embeddings (same as for other unsupervised tasks based on tabular data) into two broad types: extrinsic evaluators and intrinsic evaluators.

\noindent\textbf{Extrinsic Evaluators:} The extrinsic evaluators measure the performance of the embeddings based on downstream tasks where there may be ground truth labels available, for example in text summarization, grammar tagging, named entity recognition, sentiment analysis, etc.

\noindent\textbf{Intrinsic Evaluators:} The intrinsic evaluators directly measure syntactic or semantic relationships between words, i.e., test the quality of representations independent of specific NLP tasks. Some examples of intrinsic evaluators would be comparing similarity between words as given by, say, cosine similarity in the embedding space with similar words as perceived by humans; word analogy; concept categorization; etc.

A rigorous translation between the evaluation metrics for word embeddings to evaluation metrics for network embeddings is yet to be performed. Moreover, the corresponding datasets with ground truths that exist in the NLP domain (e.g., a list of words and their similar words as tagged by humans such as ones in Refs.~\cite{bruni2014multimodal,gerz2016simverb,hill2015simlex}) are not available in the stock networks. Hence, in the present work we take a pragmatic approach where we propose two types of evaluation metrics: a quantitative metric (or an extrinsic metric) using which we perform the hyperparmeter optimization, and a few qualitative metrics (intrinsic metrics) for which we may not have objective ground truth but may match with the common wisdom of a trader or portfolio manager.

\subsection{Quantitative Evaluation of the Embeddings}
Though there are no ground truth labels available for the stock network data that can be used to compute the embeddings for our purposes, there are industry classifications of stocks provided by the Global Industry Classification Standard (GICS) \cite{barra2009global}. GICS is a third-party provided classification system that classifies all major companies into coarse to granular categories starting from sectors (the coarsest) to industry groups, industry sub-groups, etc. (we call them GICS categories). In particular, all the S\&P 500 stocks are assigned a unique value (we call this a class in the remainder of the paper) for sector, industry group, industry sub-group, etc. GICS categories are widely used in investment processes for various purposes ranging from risk management (peer analysis), risk factor analysis, thematic investments, etc. These categories also play crucial role in mutual fund categorizations. For the purpose of the present work, we only focus on the first three levels of the classifications.

In general, stocks from the same class should be highly correlated as opposed to stocks from different classes. Hence, a \textit{good} embedding learned using Node2Vec methodology should place stocks from the same class (e.g., financial sector) closer to each other in the embedded space. In turn, if we perform clustering, for example using K-means clustering \cite{hastie2009elements}, with K being the same number as the number of classes in the chosen category, then all the K clusters should map back to the classes of the category, i.e., each cluster should only have the stocks of one and only one class. 

To evaluate the mapping between the K clusters in the given embedded space and the classes, we can employ an external entropy based cluster validation technique called \emph{V-measure}. This metric is independent of the absolute values of the labels, i.e, a permutation of the class or cluster label values does not change \emph{V-measure}. Additionally, this metric is symmetric, i.e., swapping labels with predictions returns the same score, which can be useful to measure the agreement of two independent label assignment strategies on the same dataset where the ground truth is unknown.

The \emph{V-measure} is defined as the following harmonic mean\cite{rosenberg-hirschberg-2007-v}:
\begin{displaymath}
   \emph{V-measure}=\frac{(1+\beta)(\text {Homogeneity})(\text {Completeness})}{(\beta \times \text {Homogeneity}+\text {Completeness})} 
\end{displaymath}
Here, when each of the clusters only contains data points which are members of a single ground truth class, the clustering is called homogeneous, otherwise it is called a heterogeneous cluster. Completeness is defined as the measure of a cluster when a cluster contains all the elements of a class, as opposed to a
cluster that fails to capture one or more elements of a class. $\beta$ denotes the ratio of weight attributed to homogeneity versus completeness and we use the default value of 1. 

In summary, for a learned embedding (i.e., after training the embedding algorithm at a given hyperparameter point), we first perform K-means clustering, and then compute the \emph{V-measure} between the K clusters and each of the three GICS categories. The larger the \emph{V-measure} for a hyperparameter point, the better the embedding, and the hyperparameters of Section \ref{sec:hyperparameters} algorithm were tuned to maximize \emph{V-measure}. 

\subsection{Qualitative Evaluations}
We use two qualitative evaluations methods here borrowed from word embedding evaluation methods: stock similarity and analogical inference for stocks.

\subsubsection{Stock Similarity}
In the filtered network, not all stocks are connected to all other stocks, but an embedded representation of the network is an abstract and continuous space where we can compute distance from any stock to any other stock. This means that for every stock we can identify similar stocks and rank them according to the chosen distance metric, even for the pairs of stocks which were not directly connected in the original network. Similarities from the embeddings are generated by computing pairwise cosine distance between embedding vectors for the S\&P 500 universe. The cosine distance between each pair of data-points is defined as:
\begin{displaymath}
    d^{\cos}_{i j} = 1 - \frac{\textbf{x}_i \cdot \textbf{x}_j}{\left\|\textbf{x}_i\right\|\left\|\textbf{x}_j\right\|},
    \end{displaymath}
where $\textbf{x}_i$ and $\textbf{x}_j$ are vector embeddings for the $i$-th and $j$-th stocks, respectively.

Unlike the word embedding evaluations where the modeler may have access to publicly available similarity datasets such as SimVerb-3500 \cite{gerz2016simverb} or MEN \cite{bruni2014multimodal} (although, strictly speaking, these lists may also have subjectivity built-in), there does not exist such standards for the S\&P 500 dataset. Hence, in the present paper, we rely on qualitative assessment of the list of similar stocks.

\subsubsection{Analogical Inference}
In NLP, once a word embedding is learned from a text corpus, in addition to the word similarity, one can also perform analogical inferences \cite{mikolov2013linguistic,levy2014linguistic,bolukbasi2016man} within the embedding space that captures arithmetic relationships between words. Here, given a set of three words, a, b and c, the task is to identify such word d for which the relation `c is similar to d' is the same as the relation `a is similar to b'. For instance, once a word embedding is obtained, the classic example of "Man is to King as Woman is to ?" may be solved by performing arithmetic operations on the word vectors as \emph{vector}("King")-\emph{vector}("Man")+\emph{vector}("Woman") to produce a vector representation of the word "Queen". Eventually, the results for such queries are evaluated by human experts. There are a few publicly available benchmark datasets \cite{gao2014wordrep,gladkova2016analogy,mikolov2013linguistic, mikolov2013distributed} which provide some ground truths to such problems in the NLP area.

For embedded representation learned for the stock network data, such operations between stock vectors can also be used for making analogical inferences, for example, "JPM is to GS as JNJ is to ?" produces a vector that is close to the vector representation of AMGN. Here again, there does not exist any benchmark analogical inference related datasets for stocks, and for the purpose of the present work we rely on qualitative evaluation of the results.

\section{Experiments and Results}
In this Section, we present our results from our experiments following the methodology described in the previous Section. We start by describing results from quantitative evaluation from which we identify the optimal hyperparameter point. We also show a visualization of the embedded space in passing. Then, we provide results of various qualitative evaluations.

\subsection{Quantitative Evaluation}
Table \ref{tab:hyperparameter_opt} shows the results for hyperparameter scan over various values for $l$, $r$, $p$, $q$, $w$ and dim. For each hyperparameter point, we learned the corresponding embedded space and then ran K-means clustering where K$= 9, 59$ and $169$, for GICS Industry Sector, Industry Group and Industry Sub-group, 
respectively, corresponding to as many classes in each of the GICS categories. In the Table, we record the \emph{V-measure} for each of the categories in separate columns, and the average of the \emph{V-measures} across the three categories.

As the GICS classification becomes more granular, the \emph{V-measure} increases, which means that the stocks tend to be more closely related among each other. The average of the \emph{V-measures} over all three categories is used as a tie-breaker to identify the optimal hyperparameter point, meaning that we prefer an embedding in which on average all levels of GICS classification are clustered well. The emboldened hyperparameter point in the table is chosen as the optimal point and the corresponding embedding is used in the remaining computations.

\begin{table}[htbp]
    \centering
    \resizebox{\columnwidth}{!}{\begin{tabular}{|c|c|c|c|c|c|c|c|c|c|}
    \hline
        \bfseries $l$ & \bfseries $r$ & \bfseries $p$ & \bfseries $q$ & \bfseries $w$ & \bfseries dim & \bfseries{Sector} & \bfseries{Group} & \bfseries{Subgroup} & \bfseries{Average}\ \\ \hline
        50 & 10 & 0.5 & 2 & 5 & 16 & 0.28 & 0.67 & 0.85 & 0.60  \\ \hline
        100 & 10 & 0.5 & 2 & 5 & 16 & 0.3 & 0.68 & 0.85 & 0.61  \\ \hline
        100 & 10 & 0.5 & 2 & 5 & 32 & 0.25 & 0.68 & 0.85 & 0.59  \\ \hline
        100 & 10 & 0.5 & 2 & 5 & 64 & 0.29 & 0.68 & 0.85 & 0.61  \\ \hline
        100 & 10 & 0.5 & 2 & 5 & 128 & 0.24 & 0.69 & 0.85 & 0.59  \\ \hline
        100 & 10 & 0.5 & 2 & 5 & 16 & 0.26 & 0.58 & 0.77 & 0.54  \\ \hline
        100 & 50 & 0.5 & 0.2 & 5 & 16 & 0.32 & 0.68 & 0.84 & 0.61  \\ \hline
        100 & 50 & 0.5 & 2 & 5 & 16 & 0.32 & 0.68 & 0.84 & 0.61  \\ \hline
        \textbf{100} & \textbf{50} & \textbf{2} & \textbf{0.5} & \textbf{5} & \textbf{16} & \textbf{0.35} & \textbf{0.7} & \textbf{0.84} & \textbf{0.63}  \\ \hline
        100 & 50 & 2 & 0.5 & 10 & 16 & 0.34 & 0.69 & 0.85 & 0.63  \\ \hline
        100 & 50 & 2 & 0.5 & 20 & 16 & 0.35 & 0.68 & 0.85 & 0.63  \\ \hline
        100 & 100 & 2 & 0.5 & 10 & 16 & 0.32 & 0.68 & 0.85 & 0.62  \\ \hline
        100 & 100 & 2 & 0.5 & 20 & 16 & 0.35 & 0.69 & 0.85 & 0.63  \\ \hline
        100 & 100 & 2 & 0.5 & 5 & 16 & 0.34 & 0.68 & 0.84 & 0.62  \\ \hline
        100 & 100 & 0.5 & 2 & 5 & 16 & 0.26 & 0.67 & 0.85 & 0.59  \\ \hline
        200 & 10 & 0.5 & 2 & 5 & 32 & 0.24 & 0.68 & 0.85 & 0.59  \\ \hline
        200 & 10 & 2 & 0.5 & 5 & 32 & 0.35 & 0.67 & 0.84 & 0.62  \\ \hline
        200 & 50 & 2 & 0.5 & 5 & 32 & 0.31 & 0.69 & 0.84 & 0.61  \\ \hline
    \end{tabular}
    }
    \caption{Effect of various hyperparameter combinations on \emph{V-measure} value across GICS Industry sector, group and subgroup categories as labels.}
\label{tab:hyperparameter_opt}
\end{table}

\subsection{Visualization of Embeddings}
\begin{figure}[ht]
    \centering
    \includegraphics[width=8cm, height=7cm]{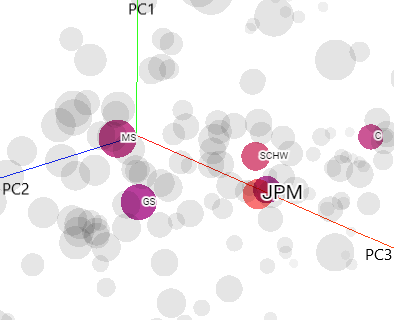}
    \caption{A 3-dimensional visualization of the 16-dimensional embedding. The plot shows a specific stock and its nearest neighbors in the lower dimensional embedding.}
    \label{fig:visualization}
\end{figure}

Figure \ref{fig:visualization} shows a 3-dimensional visualization, using the dimensionality reduction technique called principal component analysis (PCA), of the 16-dimensional stock embeddings. Here, we plot the query stock JPM and its nearest stocks in the 16-dimensional embedding space.

\subsection{Qualitative Evaluation}
We discuss results for the two qualitative evaluations: stock similarity and analogical inference for stocks.
\subsubsection{Stock Similarity}
Table \ref{tab:stock_similarity} shows the results for top 10 stocks most similar to JPM sorted according to their cosine similarity scores in the embedded space. First, notice that all the 10 similar stocks belong to the same GICS sector as JPM even in cases where there were no direct links between them in the original filtered network, i.e., they may not be directly correlated. All the 10 stocks indeed are those of either another bank or an asset manager. Upon checking many other examples, qualitatively the embedding indeed provides similarities from financial domain point of view.

\begin{table}[ht]
\centering
\resizebox{\columnwidth}{!}{\begin{tabular}{|c|c|c|}
    \hline
    \bfseries Most Similar & \bfseries Similarity Score  &  \bfseries Industry Sector \\ \hline
    GS & 0.928 & Financial \\ \hline
    BAC & 0.925 & Financial \\ \hline
    MS & 0.901 & Financial \\ \hline
    C & 0.897 & Financial \\ \hline
    SCHW & 0.857 & Financial \\ \hline
    TFC & 0.806 & Financial \\ \hline
    RJF & 0.769 & Financial \\ \hline
    USB & 0.729 & Financial \\ \hline
    NTRS & 0.725 & Financial \\ \hline
    WFC & 0.666 & Financial \\ \hline
\end{tabular}}
\caption{Top 10 most similar stocks to JPM from the embedding.}\label{tab:stock_similarity}
\end{table}

\subsubsection{Analogical Inference of Stocks}
In Table \ref{tab:analogy_inf_1}, we provide examples of analogical inference for domain experts to qualitatively evaluate the embedding.

\begin{table}[ht]
    \centering
    \resizebox{\columnwidth}{!}{\begin{tabular}{|l|c|}
    \hline
    \bfseries Analogy & \bfseries Similarity \\ \hline
        JPM is to GS as MSFT is to GOOGL & 0.882  \\ \hline
        JPM is to GS as JNJ is to AMGN & 0.837  \\ \hline
    \end{tabular}}
    \caption{Analogical inferences for JPM}
    \label{tab:analogy_inf_1}
\end{table}

We go beyond the simple analogical inference and show results for more complex queries in Table \ref{tab:analogy_inf_2}. Here, we have a set of four stocks - JPM, MS, GS and GOOGL where GOOGL does not match with the remaining three stocks as JPM, MS and GS belong to the financial sector whereas GOOGL belongs to technology sector as per the GICS Sector classification. The embeddings could make the distinction between stocks in different sectors, although this information was not explicitly supplied to the algorithm. In the second example, JNJ, BMY and PFE belong to pharmaceutical sector whereas HD belongs to consumer cyclical. Last example shows TSLA to be the most dissimilar out of the given list which aligns with intuition as TSLA belongs to consumer cyclical sector whereas UAL, AAL and DAL belong to the consumer discretionary sector.

\begin{table}[ht]
    \centering
    \resizebox{\columnwidth}{!}{\begin{tabular}{|l|c|}
    \hline
    \bfseries Analogy & \bfseries Does Not Match \\ \hline
        Does not match from JPM, MS, GS, GOOGL & GOOGL  \\ \hline
        Does not match from JNJ, BMY, PFE, HD & HD  \\ \hline
        Does not match from UAL, AAL, DAL, TSLA & TSLA  \\ \hline
    \end{tabular}}
    \caption{Analogical inference to identify the stock that does not belong to a given set.}\label{tab:analogy_inf_2}
\end{table}

Similarly, for example in Table \ref{tab:analogy_inf_3}, the embedding identifies FB as the most similar stock to GOOGL from a given set of stocks which includes JNJ, MS, MOS and FB, which aligns with intuition as JNJ belongs to the pharmaceutical sector, MS belongs to financial sector and MOS belongs to industrials sector.

\begin{table}[ht]
    \centering
    \resizebox{\columnwidth}{!}{\begin{tabular}{|l|c|}
    \hline
    \bfseries Analogy & \bfseries Similar Stock \\ \hline
        Most similar to GOOGL given JNJ, MS, MOS, FB & FB  \\ \hline
        Most similar to BLK given TSLA, STT, JNJ, AAPL & STT  \\ \hline
        Most similar to WMT given CVS, COST, JNJ, MSFT & COST  \\ \hline
    \end{tabular}}
    \caption{Analogical Inference to identify the most similar stock from a set of stocks.} 
    \label{tab:analogy_inf_3}
\end{table}



\section{Discussion and Conclusion}
The correlation matrix for stock returns data has been one of the most extensively studied objects in finance. From the network science point of view, the correlation matrix is usually transferred to a network where stocks are treated as nodes and the correlations between a pair of stocks are treated as edges. In the present work, after applying MST algorithm to sparsify the network, we use the Node2Vec algorithm to generate sentence-like structures from the network by generating random walks of chosen length from each node. Then, we applied a word embedding algorithm called Word2Vec to compute an embedded representation of the network.

We proposed an extrinsic (quantitative) evaluator based on GICS Classification of companies to evaluate the embedded representations and performed hyperparameter optimization to obtain a 16-dimensional representation of the network data. In turn, the algorithm is instructed to learn the embedding that \textit{anchors} the definition of similarity on the underlying definition of GICS classifications. In other words, the learned manifold is not an arbitrary representation of the network, rather has implicitly taken the definition of similarity as used by the GICS classification into account: for a given stock under investigation, all the stocks within its GICS class are similar to each other though there is no ranking (i.e., which of the stocks in the class is more similar than the others) provided by the classification system. The embedding, on the other hand, provides such a ranking which is extracted from GICS labels.

We then evaluated the embedding at the optimal hyperparameter point using qualitative metrics such as stock similarity and analogical inference. The stock embeddings can be used in several downstream tasks such as for building stock recommender systems, performing analogical inferences, feature creation in link prediction or node classification tasks in graph neural networks, etc. The stock embedding is essentially a compressed 16-dimensional representation of the $504\times 504$ size correlation matrix which can have multiple applications in portfolio construction as well as risk management.

Another important application of the stock embedding which we plan to explore further in the future is to use them as input features to another ML model: traditionally, nominal categorical variables are encoded as dummy variables, also known as one-hot encoded vectors, which in the case of a stock universe will have a high cardinality and hence drastically increases the dimensionality of the feature space. Additionally, one-hot encoded representation of stocks treats stocks as independent of one another and does not take into consideration the interactions that may exist between them. For example, similarity of stocks based on industry or sector classifications and pair of stocks in one-hot encoded representations are separated by a distance of $\sqrt{2}$ in the Euclidean space. Instead, using the continuous stock embeddings as features can help ML models with small datasets (i.e. even if a stock ticker is not present at the time of training the model, it will be able to understand this ticker from its embeddings).

Finally, as an example, we have analyzed stock correlation networks, but the present framework can also be applied to returns correlation matrix for any other financial assets such as mutual funds \cite{satone2021fund2vec}, hedge funds, corporate bonds, municipal bonds, etc.



\begin{acks}
The views expressed here are those of the authors alone and not of BlackRock, Inc. This document is a product of academic curiosity of the authors, and not intended as investment research or investment advice, or a recommendation, offer or solicitation for the purchase or sale of any security, financial instrument, financial product or service, or to be used in any way for evaluating the merits of participating in any transaction.
\end{acks}


\bibliography{main}{}
\bibliographystyle{unsrt}

\end{document}